      [Venezia 2006 - Il Nuovo Cimento]
\documentclass{cimento}

\usepackage{url}
\usepackage{graphicx}
\def\ltsima{$\; \buildrel < \over \sim \;$}
\def\lsim{\lower.5ex\hbox{\ltsima}}
\def\gtsima{$\; \buildrel > \over \sim \;$}
\def\gsim{\lower.5ex\hbox{\gtsima}} 

\title{REM observations of GRB\,060418: the fireball Lorentz factor determination}
\author{S.~D.~Vergani\from{ins1}\ETC ~on behalf of the REM collaboration 
}
\instlist{\inst{ins1} DIAS - DCU,
       Ireland}
\PACSes{\PACSit{98.70.Rz}{gamma-ray bursts.}}
\begin{document}

\maketitle

\begin{abstract}
We measured the fireball Lorentz factor of GRB\,060418 by the direct observation of the onset of the NIR afterglow carried with the REM telescope. We found $\Gamma_0\sim400$.
\end{abstract}

\section{Data}

REM (Rapid Eye Mount; \url{http://www.rem.inaf.it/}) is a 60 cm diameter fast reacting robotic telescope located at the ESO-La Silla in Chile, primarily designed to follow the early phases of the afterglow of GRBs detected by spaceborne $\gamma$-alert systems such as \textit{Swift} \cite{Zerbi2001,Chincarini2003}. The telescope hosts two instruments: REMIR, an infrared ($z', J, H, K'$) imaging camera, and ROSS, a visibile imager (V, R, I filters) and slitless spectrograph.

\begin{figure}
	\centering
	\includegraphics[width=7cm]{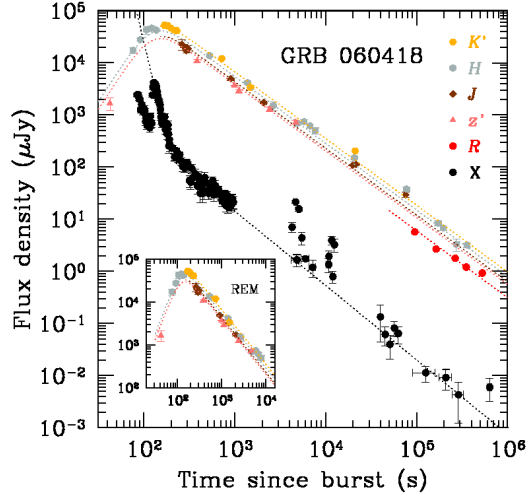}
	\caption{NIR, optical and X-ray afterglow light-curves of GRB\,060418.}
	\label{fig:060418}
\end{figure}

GRB\,060418 was detected by Swift at 03:06:08\,UT \cite{Falcone2006a}, with $T_{90} = 52 \pm 1$\,s (90\% error). The REM telescope began observing the field of GRB\,060418 64\,s after the burst (39\,s after the reception of the alert). A bright NIR source was identified \cite{Covino2006b}. REM followed the event down to the sensitivity limits in $z'JHK'$-bands. Later TNG and VLT observations were obtained by our team. The complete light-curves, including also the $z'$ point reported by \cite{GCN}, are shown in Fig.\,\ref{fig:060418} together with the \textit{Swift}-XRT light-curve.
The light-curve in the NIR band shows a rapid increase followed by a maximum and then the beginning of the regular power-law decay. As the decay after the peak is not different from the later afterglow power law, we interpret the rise as the beginning of the afterglow \cite{REM2007}.
In the X-ray band a prominent flare overimposed to the decaying X-ray afterglow, also visible in the BAT data, was observed by XRT at about 128\,s after the trigger \cite{Falcone2006b}. We interpret this component as some late activity of the inner engine, not correlated to the standard afterglow emission. 

\section{Determination of the Lorentz factor $\Gamma_0$ }
The afterglow NIR light curve of the burst peak at a time $t_{\rm peak}$ of $153$\,s, therefore $t_{\rm peak}>T_{90}$ as expected in an impulsive regime outflow 'thin shell' case \cite{Mes2006,Sari1999b}. 
The peak of the afterglow represents the time at which the dissipated power of the fireball is maximum and it can be considered coincident at first approximation with the fireball deceleration timescale $t_{\rm dec}=r_{\rm dec}/(2c\Gamma^2)$, where $r_{\rm dec}$ is the deceleration radius given by 

\begin{eqnletter}
\label{rdec}
r_{dec}=\left( \frac{3E}{4\pi~n_0m_pc^2\Gamma^2}\right) ^{1/3}
\end{eqnletter}
for an outflow total energy $E$ and an external medium particle density $n_0$.
It is therefore possible using the two relations to estimate the fireball Lorentz factor $\Gamma$ at $t_{\rm peak}/(1+z)\simeq t_{\rm dec}$ \cite{Sari1999b} which is expected to be $\Gamma\sim(1/2)\Gamma_0$ \cite{Mes2006} where $\Gamma_0$ is the initial fireball Lorentz factor.
%
Using $E_{\gamma}=9\times10^{52}$\,erg \cite{Golenetskii2006} and $z=1.489$ \cite{Dupree2006}, we obtain $\Gamma\sim200$  and $\Gamma_0\sim400$ \cite{REM2007}. This result is rather insensitive to the total energy and to the density of the external
medium and it is in agreement with the fireball model prediction of $\Gamma_0\gsim100$. This is the largest bulk Lorentz factor ever measured for any
astrophysical accreting object.

\end{document}